\documentclass{JAC2000}

\usepackage{graphicx}

\begin{document}
\title{Superconducting Superstructure for the TESLA Collider: New Results}

\author{N. Baboi, M. Liepe, J. Sekutowicz\\ Deutsches Elektronen-Synchrotron DESY, D-22603 Hamburg, Germany,\\ M. Ferrario, INFN, Frascati, Italy}

\maketitle

\begin{abstract} 
A new cavity-chain layout has been proposed for the main linac of the TESLA linear collider \cite{SUSU}. This superstructure-layout is based upon four 7-cell superconducting standing-wave cavities, coupled by short beam pipes. The main advantages of the superstructure are an increase in the active accelerating length in TESLA and a saving in rf components, especially power couplers, as compared to the present 9-cell cavities. The proposed scheme allows to handle the field-flatness tuning and the HOM damping at sub-unit level, in contrast to standard multi-cell cavities. 
The superstructure-layout is extensively studied at DESY since 1999. Computations have been performed for the rf properties of the cavity-chain, the bunch-to-bunch energy spread and multibunch dynamics. A copper model of the superstructure has been built in order to compare with the simulations and for testing the field-profile tuning and the HOM damping scheme. A "proof of principle" niobium prototype of the superstructure is now under construction and will be tested with beam at the TESLA Test Facility in 2001. In this paper we present latest results of these investigations. 
\end{abstract}

\section{INTRODUCTION}
The cost for a superconducting linear collider can be significantly reduced by minimizing the number of microwave components, and increasing the fill factor in a machine. Here the fill factor is meant as a ratio of the active cavity length to the total cavity length (active length plus interconnection).  These two conditions become partially fulfilled when the number of cells ($N$) in a structure -fed by one fundamental mode (FM) coupler- increases. Unfortunately there are two limitations on the cell's number in one accelerating structure: firstly the field flatness -the sensitivity of the field pattern increases proportional to $N^2$- and secondly trapped higher order modes (HOM). In order to overcome these limitations on $N$, the concept of the superstructure has been proposed for the TESLA main linac \cite{SUSU}. In this concept four 7-cell cavities (sub-units) are coupled by short beam tubes. The whole chain can be fed by one FM coupler attached at one end beam tube. The length of the interconnections between the cavities is chosen to be half of the wave length. Therefore the $\pi$-0 mode ($\pi$ cell-to-cell phase advance and 0 cavity-to-cavity phase advance) can be used for acceleration. In the proposed scheme HOM couplers can be attached to interconnections and to end beam tubes. All sub-units are equipped with a tuner. Accordingly the field flatness and the HOM damping can be still handled at the 7-cell sub-unit level.

\section{REFILLING OF CELLS AND BUNCH-TO-BUNCH ENERGY SPREAD}

\begin{figure}[htb]
\centering
\includegraphics*[width=65mm]{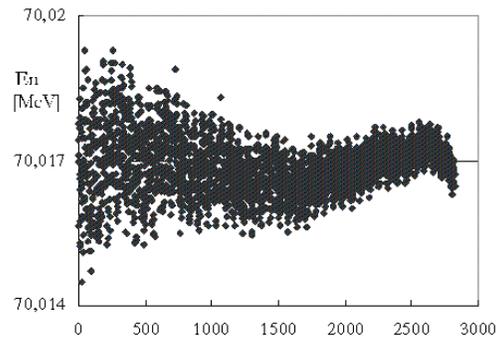}
\caption{Calculated energy gain for 2820 bunches accelerated by the proposed superstructure.}
\label{spread}
\end{figure}

The energy flow through cell-interconnections and the resulting bunch-to-bunch energy spread has been extensively studied for the superstructure with two independent codes: HOMDYN \cite{Ferrario} and  MAFIA \cite{MAFIA}\cite{Dohlus}. Negligible spread in the energy gain, smaller than $6 \cdot 10^{-5} $ for the whole train of 2820 bunches, proofs that energy flow is big enough to re-fill cells in the time between two sequential bunches; see Fig.~\ref{spread}. The energy spread results from the interference of the accelerating mode with other modes from the FM passband. The difference in energy becomes smaller at the end of the pulse due to the decay of the interfering modes.

\section{FIELD FLATNESS TUNING}
\begin{figure}[htb]
\centering
\includegraphics*[width=65mm]{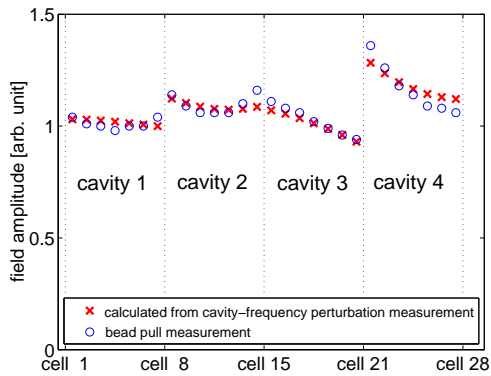}
\caption{Field profile before field flatness tuning. Shown is a comparison between the measured field profile (bead pulling on a Cu model of the superstructure) and the field profile calculated from the measured frequency perturbations of the individual cavities.}
\label{vergleich}
\end{figure}

\begin{figure}[htb]
\centering
\includegraphics*[width=65mm]{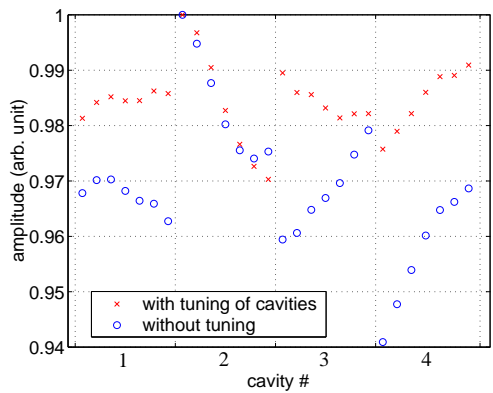}
\caption{Example of field flatness tuning by tuning the individual cavities (computer simulation). For the frequency of the individual cells a variation of $\pm$ 30 kHz is assumed.}
\label{tuning}
\end{figure}

The $\pi$-0 mode will be used for the acceleration of beam in the superstructure. Before assembly, each of the four 7-cell cavities will be pre-tuned for flat field profile and the chosen frequency of the $\pi$-0  mode. The pre-tuning procedure is based on measurements of all modes of the fundamental mode passband. It allows to adjust the profile with accuracy of better than 2-3 $\%$ for a 9-cell TESLA cavity. This error corresponds to a frequency accuracy of the individual cells of $\pm$ 30 kHz. After the cavity chain of a superstructure has been assembled and is operated in the linac at 2K, the frequency of each sub-unit can be corrected in order to equalize the mean value of the field amplitude in all sub-units (not between cells within one sub-unit). This field profile correction is possible during the linac operation, since each 7-cell structure is equipped with its own frequency tuner. The method proposed to equalize the average accelerating field of sub-units during operation is based on  perturbation theory, similar to the standard bead-pull method of L. Maier and J. Slater \cite{pert}. At first, present fields of all sub-units are measured. For that, successively, the volume of each sub-unit is changed by the same amount (stepping motor of each tuner will be moved by the same number of steps) to measure the frequency change of the  $\pi$-0  mode.  The change is proportional to the stored energy in the sub-unit of the superstructure. For each sub-unit relative values can be defined and used to calculate frequency corrections needed to equalize the field. This method has been tested on a room temperature Cu model of the superstructure -see Fig.~\ref{vergleich}- and by computer simulations, see Fig.~\ref{tuning}. One should note that the method requires only one pickup probe for all 28  cells, and therefore effectively reduces the numbers of cables, feedthroughs and electronics needed for the control. 

\section{STATISTICS OF FIELD FLATNESS}
\begin{figure}[htb]
\centering
\includegraphics*[width=65mm]{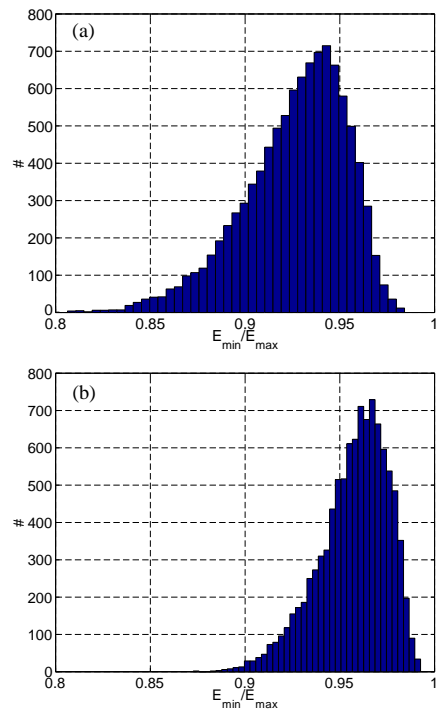}
\caption{Calculated field flatness statistics of 10000 superstructures before (a) and after field flatness tuning by adjusting the frequencies of the individual cavities (b). The frequencies of the individual cells varies by $\pm$ 30 kHz.}
\label{statistic}
\end{figure}

As discussed above the field flatness in a cold superstructure can be handled at the 7-cell sub-unit level by adjusting the frequency of each sub-unit. In order to verify this, the field flatness in a superstructure has been calculated before and after tuning of the individual cavities. The frequencies of the cavities have been corrected accordingly to the proposed tuning method. For the frequency of the individual cells a variation of $\pm$ 30 kHz is assumed, based on the experience with the TESLA 9-cell cavities. The statistics of 10000 calculated field profiles is shown in Fig.~\ref{statistic}. By adjusting the frequencies of the individual cavities the field unflatness is significantly reduced. 

\section{HOM DAMPING AND MULTIBUNCH EMITTANCE}
\begin{figure*}[htb]
\centering
\includegraphics*[width=165mm]{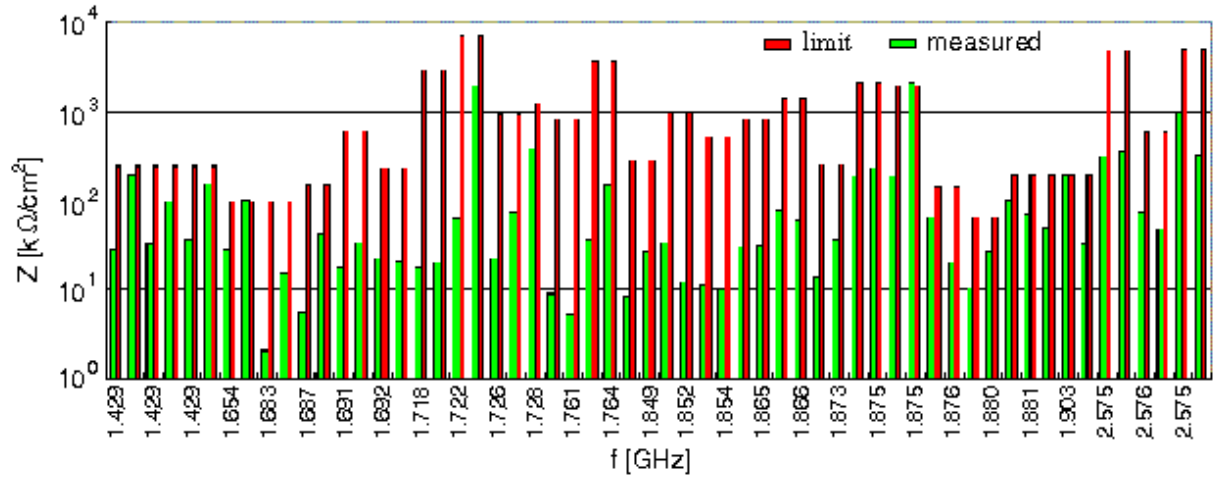}
\caption{Measured impedance values for dipole modes with higher R/Q. The impedances have been measured on a Cu model of the superstructure. For comparison also the limit is shown, based on beam dynamics simulation. }
\label{hom}
\end{figure*}

The vertical normalized multibunch emittance at the interaction point of the TESLA collider is desired to be $3 \cdot 10^{-8}$ m$\cdot$rad. Simulations of the emittance growth along the TESLA linac showed, that the dipole modes with dominating impedance (R/Q) should for that be damped to the level of $Q_{ext}<2\cdot 10^5$ \cite{EPAC2k}. The interconnecting tubes of the superstructure allow to put HOM couplers between the 7-cell cavities. Measurements on a Cu model of the superstructure at room temperature have demonstrated, that the required damping can be achieved with five HOM couplers: three attached at the interconnections and one at both ends \cite{CUSUSU}; see Fig.~\ref{hom}. Note that the sum of all listed dipole modes impedances is almost ten times smaller than the BBU limit.

\section{NB PROTOTYPE}
A first "proof of principle" niobium prototype of the superstructure is under construction \cite{NBSUSU}. The sub-units are under fabrication and will be vertical tested similar to TESLA 9-cell cavities. The beam test for the prototype is scheduled for Spring 2001. It will allow to verify  energy spread computations and RF measurements on the room temperature models. This will include the test of the HOM damping, the performance of the HOM couplers at higher magnetic field and the tuning method during operation at 2K. 

\section{CONCLUSIONS}
The presented measurements and calculations demonstrate, that in the proposed superstructure the refilling of cells, the HOM damping, the field flatness and the field flatness tuning can be handled. For the final prove, that the superstructure layout can be used for acceleration, a niobium prototype will be tested with beam at the TESLA Test Facility linac.

\section{ACKNOWLEDGEMENTS}
This work has benefited greatly from discussions with the members of the TESLA collaboration.

\end{document}